\documentclass[aps,prl,twocolumn,groupedaddress]{revtex4}
\usepackage{hyperref}
\hypersetup{colorlinks=true,urlcolor=blue}
\begin{document}

%\preprint{}

\title{wikiFactor: a measure of the importance of a wiki site.}


\author{Carl McBride}
\email[]{carl@ender.quim.ucm.es}
%\thanks{}
\affiliation{
Departamento  de Qu\'{\i}mica F\'{\i}sica.
Facultad de Ciencias Qu\'{\i}micas. Universidad Complutense de Madrid.
Ciudad Universitaria 28040 Madrid, Spain.
}


\date{\today}

\begin{abstract}
A new metric is introduced, inspired by the Hirsch {\it h-index}, to measure 
the impact of a wiki site.
A table of wikiFactors is presented for a number of wikis, in particular those
oriented towards scientific topics.
The wikiFactor is defined as the number of web pages, wF,  that have had $\ge 1000$wF visits.

\end{abstract}

%\keywords{}

\maketitle

\section{Introduction}
The Internet started as being `read-only', but has progressed  to being a `read-write'
medium, where users now have the ability to modify (some of) the web content they encounter.
This sea-change has been denoted as `web 2.0' by Tim O'Reilly.
One exemplary piece of web 2.0 software is that of the wiki, first developed
by Ward Cunningham in 1994-1995. 
This  new genre of software caught the public  eye with the appearance of 
Wikipedia in 2001. At the time of writing, the English language version of 
Wikipedia plays host to nearly three million pages of on-line user-changeable content.

It is well known that many of the defining contributions of the information technology age
have arisen from the work of scientists, scientific laboratories, or related endeavours. 
A shining  example is  the invention of the Internet by Tim Berners-Lee whilst working at CERN.
However, it is surprising to note that the wiki phenomena has, 
so far, received very little uptake amongst scientists \cite{N_2005_438_548,register},
although sooner or later this situation will have to change \cite{perish}.

In this paper a new metric is proposed to measure the importance of a wiki site. 
This metric is based on the Hirsch h-index.
\section{The Hirsch h-index}
The Hirsch h-index \cite{PNAS_2005_102_16569} is defined as the number of papers, h,
that have $\ge$h citations. This simple metric has proved to be highly popular. 
For example, it is now included as one of the functions in the ISI Web of Science. 
\section{wikiFactor}
The new metric proposed in this paper is that of the wikiFactor (wF). It 
is based on precisely the same style of measure as the h-index, but  with two differences;
the first is that it examines hits on web pages rather than citations 
of a publication, and secondly, there is a factor of a thousand. In other words 
the wikiFactor is defined as the number of pages that have had $\ge 1000$wF hits.
\subsection{Worked example}
The wikiFactor of a wiki site can be calculated in a matter of seconds. 
For example, a wikiFactor of 9 implies that the ninth most popular page has received 9,000 or more
hits or visits, whereas the tenth most popular page has not yet reached  10,000 visits.  
In Table I the wikiFactor has been calculated for a number of science-oriented
wikis. 
\begin{table}[h]%[H] add [H] placement to break table across pages
 \caption{\label{} A sample (hyperlinked) list of science oriented wiki that have a wikiFactor$>1$.}
 \begin{ruledtabular}
 \begin{tabular}{llr}
Wiki name& field & wikiFactor \\
\hline
\htmladdnormallink{OpenWetWare}{http://openwetware.org/wiki/Main_Page} & biology &  \htmladdnormallink{25}{http://openwetware.org/wiki/Special:PopularPages}  \\
\htmladdnormallink{Qwiki}{http://qwiki.stanford.edu/wiki/Main_Page} &  quantum physics & \htmladdnormallink{17}{http://qwiki.stanford.edu/wiki/Special:Popularpages}  \\
\htmladdnormallink{Folding@Home Wiki}{http://fahwiki.net/index.php/Main_Page} &  proteins & \htmladdnormallink{14}{http://fahwiki.net/index.php/Special:Popularpages}  \\
\htmladdnormallink{Quantiki}{http://www.quantiki.org/} & quantum physics &  \htmladdnormallink{12}{http://www.quantiki.org/wiki/index.php/Special:PopularPages}  \\
\htmladdnormallink{Plastics Wiki}{http://plastics.inwiki.org/Main_Page} & plastics & \htmladdnormallink{9}{http://plastics.inwiki.org/Special:Popularpages}  \\
\htmladdnormallink{Mathsoc wiki}{http://www.maths.tcd.ie/~mathsoc/w/index.php?title=Main_Page} & mathematics and physics &  \htmladdnormallink{7}{http://www.maths.tcd.ie/~mathsoc/wiki/Special:Popularpages}  \\
\htmladdnormallink{SklogWiki}{http://www.SklogWiki.org/} & thermodynamics &  \htmladdnormallink{7}{http://www.sklogwiki.org/SklogWiki/index.php/Special:PopularPages}  \\
\htmladdnormallink{Plasma-Universe.com}{http://www.plasma-universe.com/index.php/Plasma-Universe.com} & plasma  & \htmladdnormallink{6}{http://www.plasma-universe.com/index.php/Special:Popularpages}  \\
\htmladdnormallink{Psychology Wiki}{http://psychology.wikia.com/wiki/Main_Page} & psychology & \htmladdnormallink{6}{http://psychology.wikia.com/wiki/Special:Mostvisitedpages}  \\
\htmladdnormallink{NMR Wiki}{http://nmrwiki.org/} & magnetic resonance &  \htmladdnormallink{4}{http://nmrwiki.org/wiki/index.php?title=Special:Popularpages}  \\
\htmladdnormallink{Proteopedia}{http://www.proteopedia.org/wiki/index.php/Main_Page} & proteins &  \htmladdnormallink{4}{http://www.proteopedia.org/wiki/index.php/Special:Popularpages}  \\
\htmladdnormallink{PDBWiki}{http://pdbwiki.org/index.php/Main_Page} & biology &  \htmladdnormallink{2}{http://pdbwiki.org/index.php/Special:Popularpages}  \\
\htmladdnormallink{The Tangent Bundle}{http://www.physics.thetangentbundle.net/wiki/Main_Page} & physics &  \htmladdnormallink{2}{http://www.physics.thetangentbundle.net/wiki/Special:Popularpages}  \\
%\htmladdnormallink{}{} &  & \htmladdnormallink{}{}  \\
 \end{tabular}
 \end{ruledtabular}
 \end{table}
%%%%%%%%%%%%%%%%%%%%%%%%%%%%%%%%%%%%%%%%%%%%%%%%%%%%%%%%%%%%%%%%%%%%%%%%%%%%%%
\begin{acknowledgments}
%\label{acknowledgements}
This work has been funded by grants: FIS2007-66079-C02-01
from the DGI (Spain), S-0505/ESP/0299 (MOSSNOHO) from the Comunidad de Madrid,
and 910570 from the Universidad Complutense de Madrid.
\end{acknowledgments}
\bibliography{wiki}
\end{document}